\documentclass[aps,pra,reprint,superscriptaddress,nofootinbib,floatfix]{revtex4-2}
\usepackage{amsmath,amssymb,bm,graphicx,booktabs,microtype}
\usepackage[colorlinks=true,linkcolor=blue,citecolor=blue,urlcolor=blue]{hyperref}
\newcommand{\E}{\mathbb E}
\newcommand{\Var}{\operatorname{Var}}
\newcommand{\Tr}{\operatorname{Tr}}

\newcommand{\nn}{\nonumber}
\newcommand{\bea}{\begin{eqnarray}}
	\newcommand{\ena}{\end{eqnarray}}

\newcommand{\ba}{\begin{array}}
	\newcommand{\ea}{\end{array}}
\newcommand{\be}{\begin{equation}}
	\newcommand{\ee}{\end{equation}}
	
\begin{document}
\title{Higher-order disorder correlations control reflected phonon counting statistics}
\author{A.~Sedrakyan}
\affiliation{Alikhanyan National Science Laboratory, Br.\ Alikhanian 2, Yerevan 0036, Armenia}
\date{17 October 2026}
\begin{abstract}
We investigate how structural correlations beyond the power spectrum affect phonon counting in a one-dimensional acoustic stack. An integer-thickness construction cancels the extensive leading backscattering term and leaves a finite-chain reflection amplitude determined by endpoint innovations. For symmetric independent innovations with unit variance and fourth moment $\mu_4$, the normalized reflected-power second moment tends to $(\mu_4+5)/4$. With vacuum environmental inputs, a two-phonon Fock input therefore gives an ensemble-averaged reflected factorial correlation $g^{(2)}=(\mu_4+5)/8$. Two explicitly constructed ensembles with identical mean and complete two-point correlation function yield $g^{(2)}=3/4$ and $9/8$, placing their counting variances on opposite sides of the Poisson value. Conditional tensor averaging also gives a general fourth-order resistance-growth coefficient for arbitrary impedance contrast, including a positive higher-scattering contribution not recovered by shifting the lowest-order spectrum. Exact enumeration confirms the fixed-$N$, small-detuning limit and shows that increasing $N$ at fixed detuning produces a crossover governed by accumulated higher-order bulk scattering rather than convergence to the endpoint formula. Further calculations test bandwidth, thickness error, propagation loss and thermal background. The result concerns disorder-averaged pulse statistics; it neither creates nonclassicality from coherent input nor changes the conditional factorial correlation of an individually resolved passive sample.
\end{abstract}
\maketitle

\section{Introduction}
Consider an elastic wave incident on a stack of two materials with different acoustic impedances. Each interface partially reflects the wave, and the reflected amplitudes interfere. Changing the layer thicknesses changes their relative phases and therefore the reflected energy. The present question concerns fluctuations between stacks: can two ensembles with the same mean thickness and the same pair correlations nevertheless have different reflected-energy fluctuations? We first formulate this classical scattering problem, illustrated in Fig.~\ref{setup}, and then relate its reflectance moments to the counting statistics of a prepared phonon pulse. Here a phonon is a quantum of the same linear elastic wave; no phonon--phonon interaction is assumed.

Acoustic quantum systems provide a setting in which interference in an elastic structure can be probed with states of well-defined excitation number. Multi-phonon Fock states have been prepared and reconstructed in bulk acoustic-wave resonators \cite{Chu2018}, and itinerant surface-acoustic-wave phonons have been used to transfer quantum states between spatially separated nodes \cite{Dumur2021}. Quantum acoustic Fano interference further demonstrates the relevance of coherent scattering to hybrid qubit--phonon architectures \cite{Kitzman2023}. These developments motivate an examination of which structural statistics determine the counting properties of reflected phonons, beyond their effect on mean transmitted power.

In a passive linear medium, the classical scattering amplitudes determine the quantum input-output transformation, while the incident state determines the conditional counting fluctuations. Quantum-noise transport through disordered optical media has been studied experimentally and theoretically \cite{Lodahl2005a,Lodahl2005b}. The distinction between quantum fluctuations at fixed disorder and fluctuations of the scattering coefficients between disorder realizations is essential. When the realization is unresolved, the observed counting distribution is a mixture of conditional distributions. Two structures with similar average power transport can then have different averaged counting statistics.

The quantum-channel aspect has close precedents in the theory of light propagation through fluctuating media. Semenov and Vogel described the influence of atmospheric transmission fluctuations on nonclassical photon statistics \cite{SV2009}, while Bohmann \emph{et al.}\ formulated moment input-output relations for higher-order nonclassicality and entanglement criteria in fluctuating-loss channels \cite{BSSV2017}. More recently, Stepanov, Goncharov, and Kiselev studied sub-Poissonian light in fluctuating thermal-loss bosonic channels and obtained a critical mean transmittance in terms of input-state parameters, thermal occupation and a parameterized transmittance variance \cite{SGK2025}. Their analysis includes squeezed states, odd optical cat states and Fock states. These results establish that a transition across the Poisson counting-variance threshold under fluctuating attenuation is not, by itself, a new quantum-channel effect.

The present work addresses the structural origin of the fluctuating channel coefficient. Rather than prescribing its variance phenomenologically, we derive the first two reflectance moments from an elastic transfer-matrix ensemble. The same complete two-point thickness correlation function is imposed on two ensembles, while their fourth innovation moments differ. The resulting endpoint relation supplies an explicit structural realization of the established fluctuating-channel counting mechanism. Our thermal-background calculation uses the same passive beam-splitter framework as Ref.~\cite{SGK2025}; the additional content is the correlation-engineered acoustic ensemble, its controlled finite-chain limit and its higher-order resistance-growth coefficient.

The role of structural correlations is particularly clear in one-dimensional transport, where special arrangements of otherwise random scatterers can produce exceptional energies or frequencies. Several related constructions must nevertheless be distinguished. In a random binary alloy studied by Dom\'inguez-Adame, G\'omez, Avakyan, Sedrakyan, and Sedrakyan, randomness is confined to one sublattice while the other contains identical host atoms \cite{DA2000}. Their analytical and numerical treatment identifies states with localization lengths exceeding a finite sample's length near the host-site energy. This sublattice mechanism demonstrates that spatial organization of disorder matters, even when substantial randomness remains. It should not be conflated with a finite interval of extended states in an infinite generic random chain.

A complementary route uses correlated blocks of scatterers. The random-dimer construction of Dunlap, Wu, and Phillips established the importance of such short-range correlations for exceptional delocalized energies \cite{DWP1990}. Sedrakyan analyzed the random dimer model through transfer matrices, obtaining critical delocalization energies, localization-length scaling, and a relation to the density of states \cite{Sedrakyan2004}. These results establish an important distinction between the existence of an exceptional transparent energy and the way resistance grows on moving away from it. The latter requires control of the relevant transfer-matrix product and its disorder average. For fluctuation questions, one must additionally distinguish the growth of averaged resistance moments from the typical localization rate obtained by averaging a logarithm; these quantities need not share the same numerical exponent or coefficient.

The extension to coupled channels introduces further structure. Sedrakyan and Ossipov investigated a correlated two-leg ladder, calculated its Landauer resistance analytically, and compared the predicted band-center delocalization with numerical Lyapunov exponents \cite{SO2004}. Their study also examined the density of states and the distribution of Wigner delay times in the open system. It therefore supplies a precedent for examining observables beyond a mean transmission coefficient near a correlated-disorder resonance. Channel coupling and the symmetry of the transfer problem are essential in such a ladder; the single-polarization acoustic stack considered below does not reproduce that multichannel problem.

The connection to controlled quantum platforms was developed by Sedrakyan, Kestner, and Das Sarma, who proposed cold-atom realizations of the random dimer model and selected $N$-leg generalizations in optical lattices \cite{SKDS2011}. They related symmetry-dependent delocalization energies to dips in the momentum distribution. This provides a direct precedent within quantum-gas physics for translating a correlated-disorder transport property into a preparation-and-measurement protocol. The observable studied here is different: we ask how unresolved structural disorder changes factorial counting moments of reflected, initially number-controlled excitations. An optical-lattice momentum signature and an acoustic counting correlation probe different aspects of the scattering ensemble.

For layered acoustic media, Sedrakyan and Sedrakyan developed an exact calculation of mean Landauer resistance using a three-dimensional representation of the transfer matrices \cite{SS1999}. Hakobyan \emph{et al.}\ subsequently investigated exceptional transmission frequencies in two-component superlattices with correlated disorder \cite{H2000}. These works provide the acoustic starting point and the transfer-matrix method used here. Extending the average to higher tensor representations retains the same linear elastic scattering model while accessing resistance fluctuations. In particular, averaging the tensor product required for a second moment is not equivalent to taking the tensor product of the averaged first-moment propagator.

A separate but related issue arises when the disorder power spectrum suppresses the lowest-order bulk backscattering rate. Izrailev and Krokhin related weak-disorder localization to two-point correlations and their spectrum \cite{IK1999}; the review by Izrailev, Krokhin, and Makarov surveys the resulting anomalous transport in low-dimensional systems with short- and long-range correlated disorder \cite{IKM2012}. Perturbative studies of correlated disorder show that higher-order scattering can restore localization and introduce dependence on higher disorder cumulants \cite{Tessieri2002}. Resistance statistics for Markov-correlated disorder \cite{OP1997} and generalized Lyapunov exponents of random-matrix products \cite{Vanneste2010} supply further methodological precedents. A vanishing lowest-order term thus does not by itself establish exact delocalization, and sensitivity to fourth-order disorder statistics is an established possibility. It also does not make finite-range filtered disorder a long-range correlated process. The question addressed here is which measurable fluctuation remains after a specified cancellation, and whether ensembles with identical complete two-point thickness correlations can be distinguished by that fluctuation.

Here we connect an explicitly solvable acoustic disorder ensemble to a measurable factorial counting statistic. We use a two-step innovation filter whose spectrum vanishes at the leading backscattering phase. At fixed cell number, the first-order interior reflection cancels and the remaining endpoint variable has an analytically known second and fourth moment. This yields a criterion for sub- or super-Poissonian \emph{ensemble-averaged} reflected counts. Two equal-spectrum ensembles provide a finite-detuning example of this distinction. We also derive the general quartic resistance-moment rate for the same filter family, separating the shifted-spectrum contribution from an irreducible higher-order term. Numerical calculations retain the actual layer phases and use reflectance moments directly rather than replacing them by functions of averaged resistance.

The acoustic model is a lossless scalar polarization at normal incidence, supplemented by explicitly stated perturbations when robustness is examined. Its quantization assumes linear elastic propagation, no frequency conversion and a specified incident pulse mode. Connections to present quantum-acoustic experiments identify the relevant preparation and detection capabilities; they do not establish a material-specific realization of the multilayer design. We therefore quantify thermal-background and counting-rate requirements alongside the ideal prediction.

\section{Acoustic scattering and the filtered thickness ensemble}
\subsection{Geometry, elastic fields and the layer transfer matrix}
\begin{figure*}[t]
\centering
\includegraphics[width=.86\textwidth]{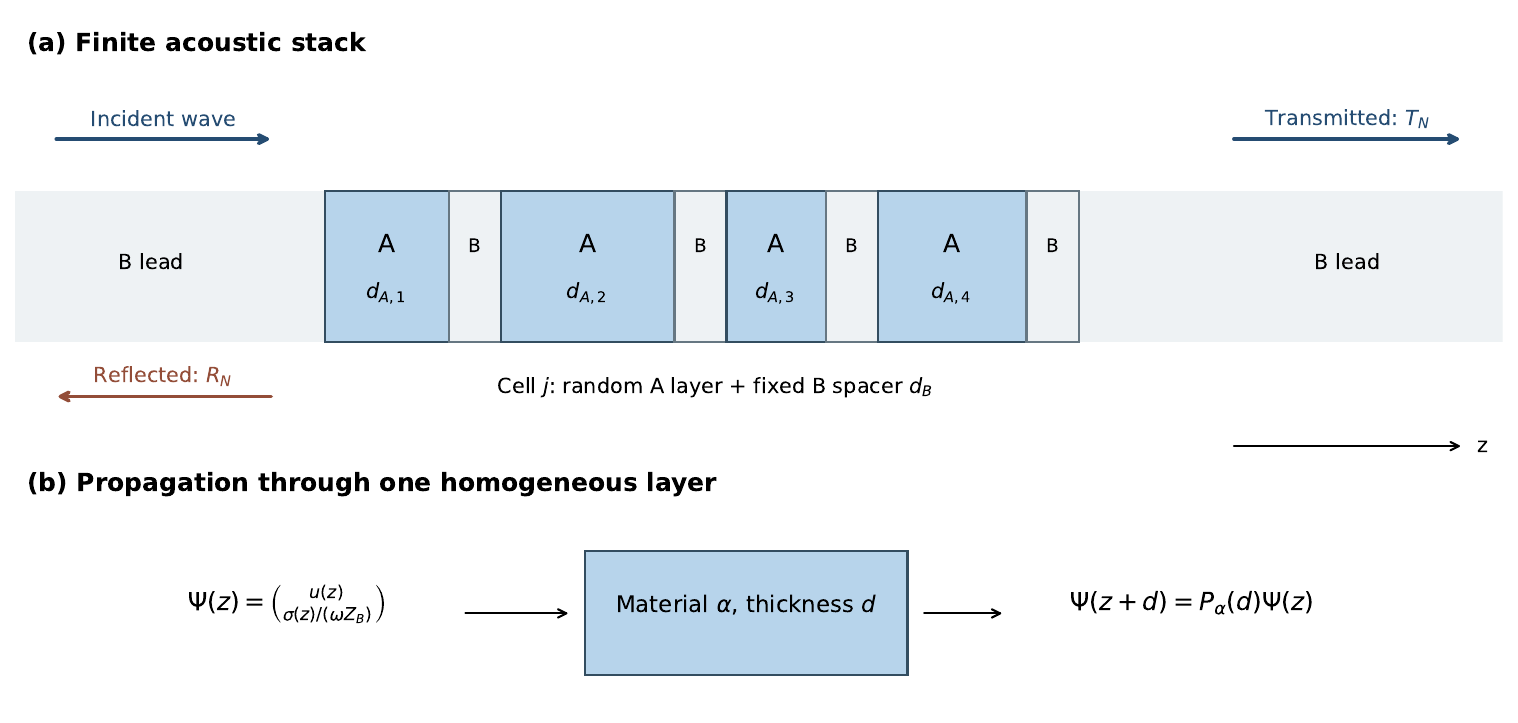}
\caption{Scattering geometry and the meaning of the layer propagator. (a) A wave enters a finite stack from a lead of material $B$. The $A$ layers have random thicknesses $d_{A,j}$, while all $B$ spacers have the same thickness $d_B$. The reflected and transmitted energy fractions are $R_N$ and $T_N$. The drawing is schematic and does not fix the thickness ratios. (b) Equation~\eqref{layer} propagates displacement and normalized stress from the left face of one homogeneous layer to its right face. Interface continuity permits these propagators to be multiplied directly.}
\label{setup}
\end{figure*}

The sample consists of $N$ cells arranged along a coordinate $z$ perpendicular to the interfaces (Fig.~\ref{setup}). Each cell contains a layer of material $A$ and a spacer of material $B$. The incoming wave travels from the left through a semi-infinite $B$ lead; the outgoing transmitted wave enters an identical $B$ lead on the right. Only the $A$-layer thicknesses fluctuate. We assume planar, perfectly bonded interfaces, linear elasticity, normal incidence and a single decoupled polarization. In this section the materials are nondispersive and have no absorption. For a longitudinal wave, the scalar field below is the displacement along $z$; the same equations describe a decoupled shear polarization with the appropriate shear modulus and speed.

Write the real displacement as $\operatorname{Re}[u(z)e^{-i\omega t}]$, where $\omega>0$ is the angular frequency and $u(z)$ is its complex amplitude. In material $\alpha\in\{A,B\}$, let $\varrho_\alpha$ be the mass density, $C_\alpha$ the relevant elastic modulus, $c_\alpha=\sqrt{C_\alpha/\varrho_\alpha}$ the wave speed, and $Z_\alpha=\varrho_\alpha c_\alpha$ the acoustic impedance. The stress amplitude is $\sigma(z)=C_\alpha\partial_z u(z)$, with tensile stress taken positive. Force balance gives
\[
 \partial_z\sigma=-\varrho_\alpha\omega^2u,\qquad
 \partial_z^2u+k_\alpha^2u=0,\qquad k_\alpha=\omega/c_\alpha.
\]
Thus the displacement is a superposition of right- and left-moving plane waves within each homogeneous layer.

To match these waves across interfaces, introduce the two-component state
\[
 \Psi(z)=\begin{pmatrix}u(z)\\ \sigma(z)/(\omega Z_B)\end{pmatrix},
 \qquad \Psi(z+d)=P_\alpha(d)\Psi(z).
\]
Both entries have the dimensions of displacement. The same reference impedance $Z_B$ is used in every layer, so continuity of displacement and stress is simply continuity of $\Psi$. Solving the wave equation over a homogeneous thickness $d$ gives
\begin{equation}
\begin{aligned}
 &P_\alpha(d)=\begin{pmatrix}
 \cos\phi_\alpha &(Z_B/Z_\alpha)\sin\phi_\alpha\\
 -(Z_\alpha/Z_B)\sin\phi_\alpha&\cos\phi_\alpha
 \end{pmatrix},\\
 &\phi_\alpha=k_\alpha d=\omega d/c_\alpha.
 \label{layer}
\end{aligned}
\end{equation}
Equation~\eqref{layer} is a spatial propagator, not a matrix of reflection and transmission amplitudes. Its phase $\phi_\alpha$ is the phase accumulated across the layer. The upper-right entry converts the initial stress into a displacement change; the lower-left entry converts the initial displacement into a stress change. The minus sign follows from the harmonic force-balance equation above. At $d=0$ the matrix is the identity, and its determinant is one. For a $B$ layer it is a rotation matrix because $Z_\alpha=Z_B$. Reflection emerges when these propagators are combined with the incoming and outgoing wave boundary conditions; a homogeneous layer with impedance different from the leads generally reflects even though its propagator has unit determinant.

\subsection{Correlated thicknesses and the reference frequency}
Let $j$ index cells, whereas $\alpha$ labels the material. The positive reference frequency $\omega_*$ fixes the designed layer thicknesses; it is not a separate excitation. The $j$th cell comprises an $A$ layer followed by a fixed $B$ spacer, with
\bea
 d_{A,j}=n_j\frac{\pi c_A}{\omega_*}, &&\quad
 n_j=n_0+\xi_j+\xi_{j-2},\\
 \frac{\omega_* d_B}{c_B} &=&\pi/4.\nn
 \label{design}
\ena
Here $n_j$ is the number of half wavelengths of material $A$ contained in layer $j$ at $\omega_*$. The integer $n_0$ sets its mean, and $\xi_j$ are auxiliary random variables (innovations) used to generate correlated thicknesses. The same innovation enters two layers separated by two cells, producing finite-range correlations. The innovations are independent, symmetric, bounded integers with
$\E\xi=0$, $\E\xi^2=1$ and $\E\xi^4=\mu_4$. Choose integer $n_0>2\max|\xi|$ to keep every layer thickness positive.

Define the detuning and impedance combinations
\begin{equation}
 \begin{aligned}\epsilon&=\frac{\omega-\omega_*}{\omega_*},\quad x=\pi\epsilon,\quad r=Z_A/Z_B,\\ \Delta&=r-r^{-1},\quad\Sigma=r+r^{-1}.\end{aligned}
\end{equation}
Up to an integer half-wave sign, which does not affect number statistics,
\begin{equation}
\begin{aligned}
 B(n_j,x)&=R(\pi/4+x/4)\\
 &\times\begin{pmatrix}\cos(n_jx)&r^{-1}\sin(n_jx)\\-r\sin(n_jx)&\cos(n_jx)
 \end{pmatrix},\\
 R(\beta)&=\begin{pmatrix}\cos\beta&\sin\beta\\-\sin\beta&\cos\beta\end{pmatrix}.
\end{aligned}
\label{block}
\end{equation}
At $x=0$, each $A$ layer is a half-wave multiple and every realization transmits perfectly between the matched leads.

Let $\delta n_j=n_j-n_0$. We use $\E$ for an average over independently generated stacks and $\Var X=\E X^2-(\E X)^2$ for their variance. The covariance is $C_\ell=\E[\delta n_j\delta n_{j+\ell}]$, where $\ell$ is an integer cell separation. Its Fourier transform is $S_n(q)=\sum_{\ell=-\infty}^{\infty}C_\ell e^{-iq\ell}$, with dimensionless wave number $q$ conjugate to the cell index. For the construction above, the thickness covariance and power spectrum are
\begin{equation}
\begin{aligned}
 &C_\ell=2\delta_{\ell,0}+\delta_{\ell,2}+\delta_{\ell,-2},\\
 &
 S_n(q)=2+2\cos(2q)=4\cos^2q.
 \label{spectrum}
\end{aligned}
\end{equation}
They do not depend on $\mu_4$. Equality here refers to the cell-index thickness spectrum; it does not assert equality of the full continuum impedance spectrum at arbitrary frequency. Since the leading reflection amplitudes from successive cells differ in phase by $2\theta=\pi/2$, the lowest-order bulk backscattering samples precisely the zero of this spectrum.

\subsection{From transfer matrices to measurable power fractions}
The cell matrix $B_j=B(n_j,x)$ propagates through the $A$ layer first and the $B$ spacer second, so the rightmost matrix acts first. The full sample propagator is $M_N=B_N\cdots B_1$. In either external lead, the state at a reference plane can be written as $\Psi=Q(a_+,a_-)^T$, where $a_+$ and $a_-$ are the displacement amplitudes of right- and left-moving waves: a right-moving wave has $\sigma/(\omega Z_B)=iu$, and a left-moving wave has $\sigma/(\omega Z_B)=-iu$. With

$Q=\left(\begin{smallmatrix}1&1\\i&-i\end{smallmatrix}\right)$, set $U_N=Q^{-1}M_NQ$. This matrix transfers traveling-wave amplitudes between the leads. For unit incident amplitude and no incidence from the right, $(t,0)^T=U_N(1,r_{\rm amp})^T$, where $r_{\rm amp}$ and $t$ are reflection and transmission amplitudes. Thus $r_{\rm amp}=-U_{21}/U_{22}$ and $t=1/U_{22}$. Because both leads are identical, the reflected and transmitted energy-flux fractions are $R_N=|r_{\rm amp}|^2$ and $T_N=|t|^2$. The dimensionless Landauer resistance $\rho_N$ measures reflected relative to transmitted power, not electrical resistance. Explicitly,
\begin{equation}
\begin{aligned}
 &R_N=\left|\frac{U_{21}}{U_{22}}\right|^2,\\
 &
 T_N=|U_{22}|^{-2},\\
 &
 \rho_N=\frac{R_N}{T_N}=\frac{\Tr(M_NM_N^T)-2}{4}.
 \label{scattering}
\end{aligned}
\end{equation}
For the lossless model $R_N+T_N=1$ and $R_N=\rho_N/(1+\rho_N)$. We calculate $R_N$ for each realization before averaging. In particular, $\E R_N$ cannot generally be reconstructed from $\E\rho_N$ alone.

\section{Endpoint cancellation and finite-chain reflectance moments}
At fixed $N$, the leading reflection amplitude is proportional to
\begin{equation}
 b_N=\frac{\Delta}{2}x\sum_{j=0}^{N-1}n_j i^j+O_N(x^2),
 \label{born}
\end{equation}
up to an irrelevant phase. Its squared modulus equals $\rho_N$; to leading order it also equals $R_N$. The filter identity
\begin{equation}
 \sum_{j=0}^{N-1}(\xi_j+\xi_{j-2})i^j
 =\xi_{-2}+i\xi_{-1}+i^{N-2}\xi_{N-2}+i^{N-1}\xi_{N-1}
 \label{telescoping}
\end{equation}
holds for $N\geq2$. It cancels all interior first-order contributions. If $N=4,8,\ldots$, the mean term $n_0\sum_j i^j$ vanishes as well. Consequently,
\begin{equation}
\begin{aligned}
 &R_N=F Y+O_N(x^3),\\
 &
 F=\frac{\Delta^2x^2}{4},\quad
 Y=(\xi_{-2}-\xi_{N-2})^2+(\xi_{-1}-\xi_{N-1})^2.
 \label{endpoint}
\end{aligned}
\end{equation}
The four innovations appearing here are independent. If $X=\xi-\xi'$, then
$\E X^2=2$ and $\E X^4=2\mu_4+6$. Therefore
\begin{align}
 \E Y&=4,&\Var Y&=4(\mu_4+1),&\E Y^2&=4\mu_4+20.
\end{align}
The leading physical reflectance moments are thus
\begin{align}
 \overline R:=\E R_N&=4F+O_N(x^3),\\
 \Var R_N&=4(\mu_4+1)F^2+O_N(x^5),\\
 C_R:=\frac{\E R_N^2}{(\E R_N)^2}
 &\longrightarrow\frac{\mu_4+5}{4}.
 \label{CR}
\end{align}
The unit-variance normalization $\E\xi^2=1$ is used throughout. More generally, for $v=\E\xi^2>0$ and $\mu_4=\E\xi^4$, the endpoint moments are $\E Y=4v$, $\Var Y=4(\mu_4+v^2)$ and $\E Y^2=4\mu_4+20v^2$. Consequently $\overline R=4vF+O_N(x^3)$ and $C_R\to(\mu_4/v^2+5)/4$. All numerical results below use $v=1$.
The normalized limit is independent of the nonzero impedance contrast; the reflected power itself vanishes as $\Delta^2x^2$. At $r=1$ there is no reflection and the normalized ratio is undefined. The limit is taken as $x\to0$ at fixed $N$ divisible by four, with the specified matched leads. It is not a large-$N$ localization limit: increasing $N$ at fixed nonzero detuning accumulates the higher-order bulk terms omitted from Eq.~\eqref{endpoint}.

A directly measurable classical counterpart is the sample-to-sample variance of reflected pulse energy. For a fixed incident energy $E_{\rm in}$ and a monochromatic or sufficiently narrowband pulse, $E_{\rm ref}=E_{\rm in}R_N$. Hence the normalized energy second moment is $C_R$. This offers a classical calibration of the structural fluctuations before a few-phonon experiment is attempted.

\section{Quantum input-output map and pulse counting}
Let $a_{\rm in}$ denote a normalized incident acoustic pulse mode. For a fixed elastic realization $s$, linear scattering transforms the reflected mode as
\begin{equation}
\begin{aligned}
 &a_{{\rm ref},s}=\sqrt{R_s}\,e^{i\varphi_s}a_{\rm in}
 +\sum_\nu c_{s\nu}v_\nu,\\
 &
 R_s+\sum_\nu|c_{s\nu}|^2=1,
 \label{quantumchannel}
\end{aligned}
\end{equation}
where the remaining input modes are initially vacuum. In a lossless two-port system the sum contains the other lead input; with absorption it also includes loss-bath operators. Those operators are required to preserve the bosonic commutator. There is no phonon interaction or parametric amplification in this model.

Let $K$ be the reflected excitation count per pulse and define the pulse-integrated factorial correlation
\begin{equation}
 g^{(2)}=\frac{\langle K(K-1)\rangle}{\langle K\rangle^2}.
 \label{gdefinition}
\end{equation}
The brackets include quantum counting and, when indicated, the ensemble of structures. They are not an average of separately normalized sample correlations. For an input state independent of $s$, with mean count $\bar n$ and factorial correlation $g_{\rm in}^{(2)}$,
\begin{align}
 \langle K\rangle_s&=R_s\bar n,\\
 \langle K(K-1)\rangle_s&=R_s^2\bar n^2g_{\rm in}^{(2)}.
\end{align}
Averaging these unnormalized moments gives the standard random-attenuation relation
\begin{equation}
 g_{\rm ref}^{(2)}=g_{\rm in}^{(2)}C_R.
 \label{quantummap}
\end{equation}
Its general physical basis is the established quantum theory of scattering in random media \cite{Lodahl2005a,Lodahl2005b}. The model-specific result follows by inserting Eq.~\eqref{CR}:
\begin{equation}
 g_{\rm ref}^{(2)}\longrightarrow
 g_{\rm in}^{(2)}\frac{\mu_4+5}{4}.
 \label{generalquantum}
\end{equation}
For a Fock input $|m\rangle$, $g_{\rm in}^{(2)}=1-1/m$. In particular,
\begin{equation}
 g_{{\rm ref},|2\rangle}^{(2)}\longrightarrow\frac{\mu_4+5}{8}.
 \label{Fock2}
\end{equation}
The second-order counting statistic is sub-Poissonian for $\mu_4<3$ and super-Poissonian for $\mu_4>3$. At $\mu_4=3$ its variance equals its mean in the limit; this does not imply that the full count distribution is Poisson.

For the two-phonon input, the complete averaged reflected count distribution is particularly simple:
\begin{align}
 P(K=2)&=\E R^2,\\
 P(K=1)&=2(\E R-\E R^2),\\
 P(K=0)&=1-2\E R+\E R^2.
 \label{countdistribution}
\end{align}
This provides an independent verification of Eq.~\eqref{quantummap}. The ordinary count variance is
$\Var K=\langle K\rangle+[g^{(2)}-1]\langle K\rangle^2$.

\subsection{What changes, and what does not}
For each individually resolved passive sample with vacuum baths, $g_{{\rm ref},s}^{(2)}=g_{\rm in}^{(2)}$ when the reflected mean is nonzero. The new dependence concerns the pooled ensemble, where random attenuation adds classical mixture fluctuations. It is not a change in the conditional bosonic statistics at fixed structure. A coherent input gives $g_{\rm ref}^{(2)}=C_R\geq1$ and cannot acquire sub-Poissonian statistics through this passive process. A super-Poissonian value for a reflected Fock input does not prove that its state is classical; it means that this second-order nonclassicality witness is no longer satisfied. A single-phonon input has $g^{(2)}=0$ for both ensembles and cannot show the proposed second-order contrast.

\section{Equal-spectrum ensembles with different counting regimes}
Choose $n_0=5$, $r=2$ and either innovation law
\begin{align}
 \mathcal A:&\quad \Pr(\xi=\pm1)=1/2,\\
 \mathcal B:&\quad \Pr(\xi=\pm2)=1/8,\quad\Pr(\xi=0)=3/4.
\end{align}
Both have unit variance and yield exactly the same covariance and spectrum in Eq.~\eqref{spectrum}. They have fourth moments one and four. The allowed $A$-layer thicknesses are positive half-wave multiples in both ensembles. Equations~\eqref{CR} and \eqref{Fock2} give
\begin{center}
\small
\setlength{\tabcolsep}{4pt}
\begin{tabular}{cccc}\toprule
Ensemble & $\mu_4$ & coherent $g_{\rm ref}^{(2)}$ & two-phonon $g_{\rm ref}^{(2)}$\\\midrule
$\mathcal A$ & 1 & $3/2$ & $3/4$\\
$\mathcal B$ & 4 & $9/4$ & $9/8$\\\bottomrule
\end{tabular}
\end{center}
These limits correspond to the same leading mean reflected power. The difference is carried by fourth-order structural statistics rather than the pair correlator.

\begin{figure*}[htbp]\centering
\includegraphics[width=.8\textwidth]{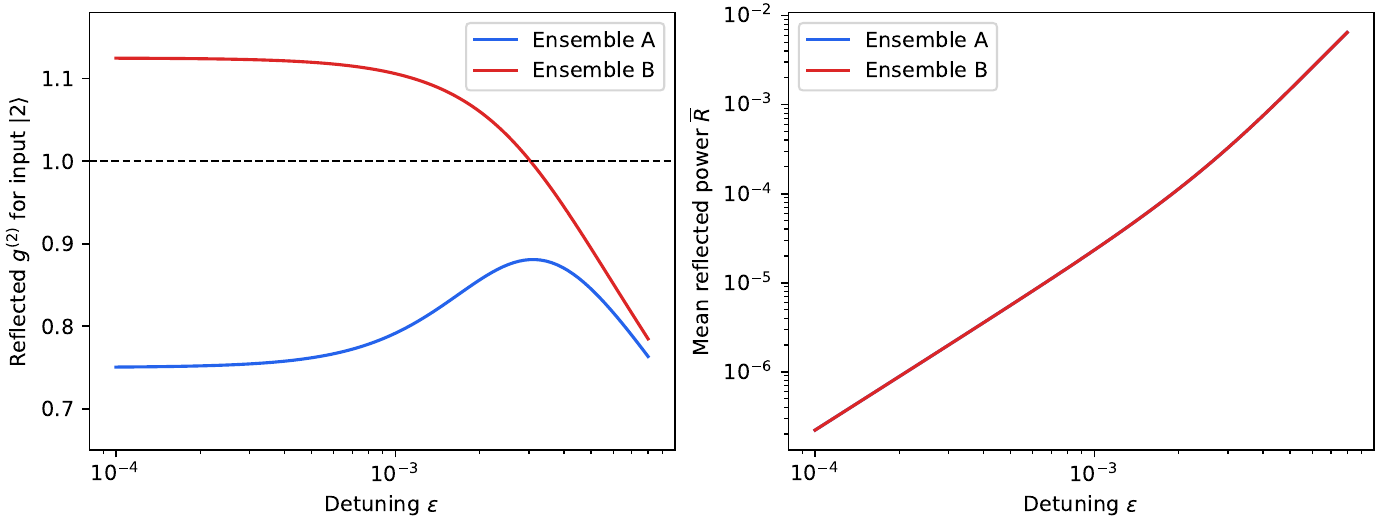}
\caption{Exact reflectance calculations for four cells, $n_0=5$ and $r=2$. Left: pooled reflected factorial correlation for input $|2\rangle$ in the two equal-spectrum ensembles; the dashed line is the Poisson counting-variance threshold. Right: their nearly identical mean reflected powers.}\label{quantumfig}
\end{figure*}

Exact weighted enumeration of every innovation word gives finite-detuning reflectance moments directly. The physical reflectance is evaluated realization by realization from
$\rho_N=[\Tr(M_NM_N^T)-2]/4$ and $R_N=\rho_N/(1+\rho_N)$ before either moment is averaged. For $N=4$ and a two-phonon input,
\begin{center}
\small
\setlength{\tabcolsep}{4pt}
\begin{tabular}{ccccc}\toprule
$\epsilon$ & $\overline R_{\mathcal A}$ & $\overline R_{\mathcal B}$ & $g_{\mathcal A}^{(2)}$ & $g_{\mathcal B}^{(2)}$\\\midrule
$0.001$ & $2.34258\,10^{-5}$ & $2.34272\,10^{-5}$ & $0.79162$ & $1.10596$\\
$0.002$ & $1.13552\,10^{-4}$ & $1.13572\,10^{-4}$ & $0.85615$ & $1.06011$\\\bottomrule
\end{tabular}
\end{center}
At the second working point the mean reflectances differ by only about $0.018\%$, while their counting variances remain on opposite sides of the Poisson value. The contrast is restricted to a finite frequency window; for example, at $\epsilon=0.005$ both ensembles are sub-Poissonian for this input. Figure~\ref{quantumfig} shows the finite-detuning behavior, rather than extrapolating the fixed-$N$ asymptotic formula indefinitely.

The order of limits can be checked directly. Table~\ref{Ndependence} lists exact values of $C_R$ for $N=4$ and $8$ while the detuning is reduced. Both ensembles approach their fixed-$N$ limits, $3/2$ and $9/4$. At fixed detuning, however, increasing $N$ need not move the result closer to those endpoint values because the $O_N(x^3)$ and higher bulk terms accumulate. For example, at $\epsilon=10^{-3}$ one finds $C_{R,\mathcal A}=1.58324,1.68443,1.69133$ and $C_{R,\mathcal B}=2.21192,2.09741,1.93485$ for $N=4,8,12$. These numbers test the finite-detuning crossover, not the fixed-$N$ limit in Eq.~\eqref{CR}. In particular, $g_{\mathcal B}^{(2)}$ crosses the Poisson value between $N=8$ and $N=12$ at this detuning, so the contrast between sub- and super-Poissonian counts established at $N=4$ is a finite-chain, small-detuning property rather than a large-$N$ one.

\begin{table}[tb]
\caption{Exact normalized reflected-power moment $C_R$ versus detuning for $n_0=5$ and $r=2$. Complete weighted enumeration is used for both innovation laws. The limiting values are $3/2$ for $\mathcal A$ and $9/4$ for $\mathcal B$.}
\label{Ndependence}
\begin{ruledtabular}
\begin{tabular}{ccccc}
$\epsilon$ & $\mathcal A$, $N=4$ & $\mathcal A$, $N=8$ & $\mathcal B$, $N=4$ & $\mathcal B$, $N=8$\\
\hline
$4.0\times10^{-3}$ & 1.74070 & 1.43122 & 1.88962 & 1.46143\\
$2.0\times10^{-3}$ & 1.71230 & 1.66802 & 2.12023 & 1.80993\\
$1.0\times10^{-3}$ & 1.58324 & 1.68443 & 2.21192 & 2.09741\\
$5.0\times10^{-4}$ & 1.52342 & 1.57572 & 2.23963 & 2.20870\\
$2.0\times10^{-4}$ & 1.50384 & 1.51407 & 2.24825 & 2.24313\\
$1.0\times10^{-4}$ & 1.50096 & 1.50358 & 2.24956 & 2.24826\\
\end{tabular}
\end{ruledtabular}
\end{table}

\section{General quartic resistance-growth coefficient}
The endpoint statistic coexists with bulk scattering at higher order. Define
\begin{equation}
 \Gamma_1=\lim_{N\to\infty}\frac1N\ln\E(1+\rho_N).
\end{equation}
To separate phase drift from higher scattering, allow the auxiliary spacer phase
$\theta_B=\pi/4+b x$. The fixed non-dispersive spacer in Eq.~\eqref{design} corresponds to $b=1/4$. The periodic cell with mean thickness $n_0$ has Bloch-phase slope
\begin{equation}
 \kappa=b+\frac{n_0\Sigma}{2}.
\end{equation}
Conditional tensor perturbation for the filter in Eq.~\eqref{design} gives the general coefficient
\begin{equation}
 \Gamma_1=\Delta^2\left[8\kappa^2+\frac{\mu_4+1}{2}\Sigma^2\right]x^4+O(x^5).
 \label{generalrate}
\end{equation}
The coefficients through cubic order vanish. The first term is reproduced by inserting the shifted periodic Bloch phase into the lowest-order spectrum formula. The second is a higher-scattering contribution and depends explicitly on the fourth moment.

Because $\mu_4\geq(\E\xi^2)^2=1$, Eq.~\eqref{generalrate} implies the positive bound
\begin{equation}
\begin{aligned}
 &\lim_{x\to0}\frac{\Gamma_1}{x^4}\geq\Delta^2\Sigma^2\\
 &(r\ne1),
 \label{floor}
\end{aligned}
\end{equation}
even if the formal phase drift is compensated by setting $b=-n_0\Sigma/2$. This is a constraint within the specified filter family and unit-variance normalization, not a no-go theorem for every correlated structure. An ordinary fixed nondispersive spacer does not have an independently tunable negative slope $b$.

For $r=2$, $n_0=5$ and $b=1/4$, Eq.~\eqref{generalrate} yields $\Gamma_{1,\mathcal A}=12393x^4/16+O(x^5)$ and $\Gamma_{1,\mathcal B}=25461x^4/32+O(x^5)$. Thus the equal-spectrum ensembles differ both in the finite-chain counting limit and in their leading nonzero bulk moment rate. This rate is a generalized moment-growth exponent, not the logarithmic typical Lyapunov exponent. The quantum counting results above require finite-chain reflectance moments and do not rely on identifying these exponents.

\section{Bandwidth, fabrication errors and passive loss}
\subsection{Pulse bandwidth}
For a normalized spectral intensity $|f(\omega)|^2$, define the reflected fraction of one pulse in each static realization as
\begin{equation}
 R_{s,f}=\int d\omega\,|f(\omega)|^2R_s(\omega).
\end{equation}
If detection collects the full reflected pulse and both input phonons occupy the same incident pulse mode, the counting relation remains Eq.~\eqref{quantummap} with $R_s$ replaced by $R_{s,f}$. It is essential to integrate each realization before squaring and averaging; averaging only the frequency-dependent second moments would discard cross-frequency correlations of the same structure.

Gaussian spectral intensities were evaluated by 24-point Gauss--Hermite quadrature. At carrier detuning $\epsilon=0.001$ and RMS fractional width $\sigma_\epsilon=0.0003$, the two counting correlations are $0.79944$ and $1.09265$. Their opposite counting regimes survive this bandwidth in the ideal vacuum model.

\subsection{Thickness errors}
Let the actual thickness be $(n_j+\eta_j)\pi c_A/\omega_*$, with independent zero-mean Gaussian $\eta_j$ of RMS $s_d$. This adds residual phase $\pi(1+\epsilon)\eta_j$. The added uncorrelated Born mean is of order $N\Delta^2\pi^2s_d^2/4$, whereas the intended endpoint mean is $\Delta^2\pi^2\epsilon^2$. A useful small-error condition is therefore
\begin{equation}
 s_d\ll\frac{2|\epsilon|}{\sqrt N}.
 \label{errorbound}
\end{equation}
This is an estimate, not a replacement for the full scattering calculation. A Monte Carlo study using 150000 words per ensemble, with 20 batches to estimate sampling errors, gives at $\epsilon=0.001$ and $s_d=0.0003$
\begin{equation}
\begin{aligned}
 &g_{\mathcal A}^{(2)}=0.8241\pm0.0014,\\
 &
 g_{\mathcal B}^{(2)}=1.0892\pm0.0026.
\end{aligned}
\end{equation}
The uncertainties describe Monte Carlo sampling only. Increasing $s_d$ progressively drives the two statistics toward one another.

\subsection{A phenomenological propagation-loss test}
As a narrowband test, replace each propagation wave number by $k(1+i\ell)$ while retaining real interface impedances. Here $\ell=\operatorname{Im}k/\operatorname{Re}k$ attenuates the full physical phase, not only its residual detuning. After removal of the real half-wave sign, the phases used in the cell matrices are
\begin{equation}
\begin{aligned}
 \widetilde\phi_{A,j}&=n_jx+i\ell n_j\pi(1+\epsilon),\\
 \widetilde\phi_B&=\frac\pi4(1+\epsilon)(1+i\ell).
\end{aligned}
\label{lossphases}
\end{equation}
In particular, removing the sign $(-1)^{n_j}$ does not remove the attenuation associated with the physical thickness $n_j\pi c_A/\omega_*$. This represents matched distributed attenuation, not a fit to a particular elastic damping mechanism. The reflected amplitudes are calculated with the complex transfer matrices; the lossless resistance identity in Eq.~\eqref{scattering} is not used. We check $R+T\leq1$ for every realization. Vacuum bath operators complete the passive quantum channel in Eq.~\eqref{quantumchannel}.

\begin{figure*}[tb]\centering
\includegraphics[width=.8\textwidth]{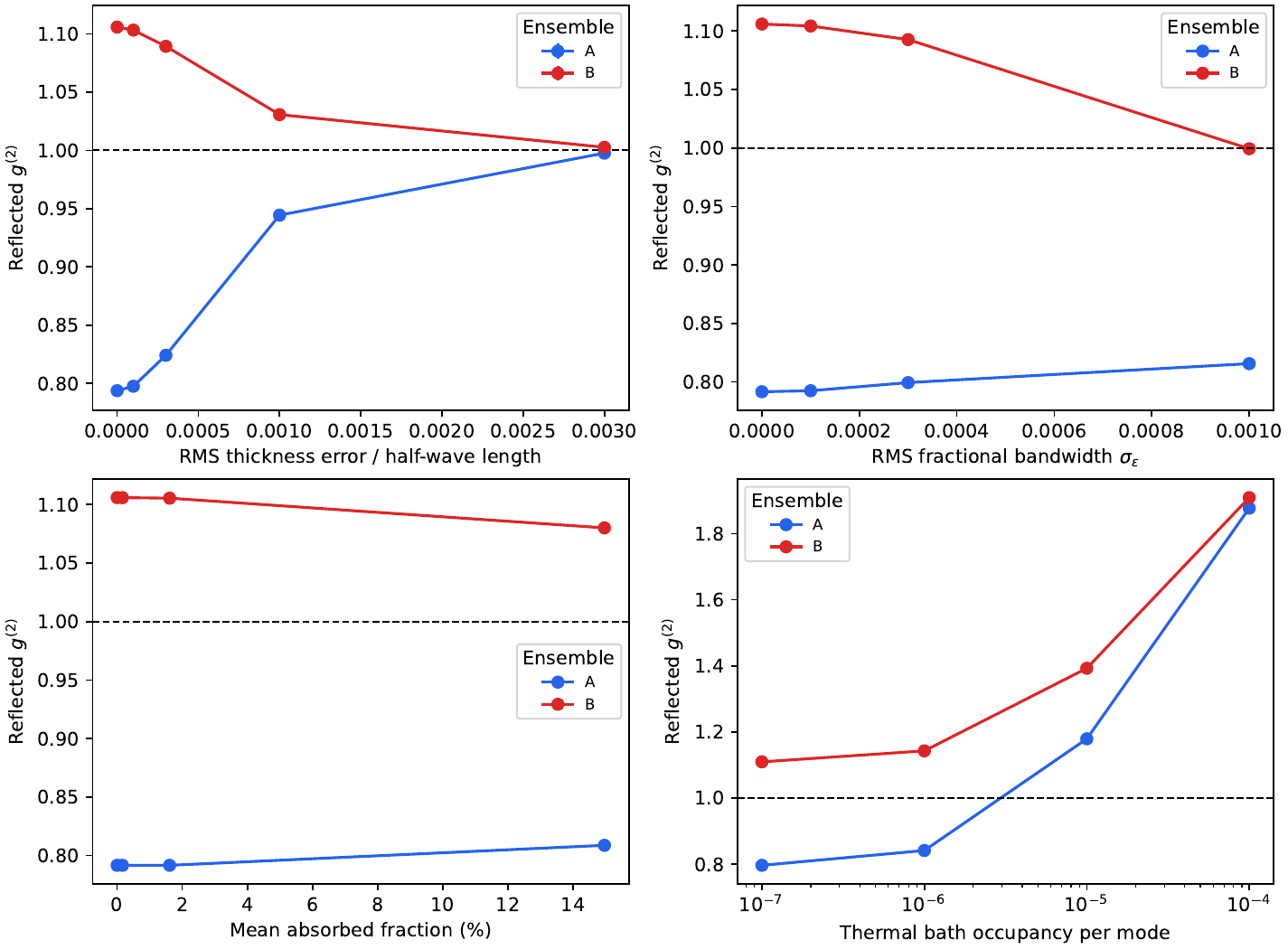}
\caption{Robustness tests for input $|2\rangle$, four cells, $r=2$ and carrier detuning $\epsilon=0.001$. Thickness-error points are Monte Carlo estimates with batch standard errors; bandwidth and propagation-loss points are numerical evaluations of the finite ensemble. Thermal-background values follow Eq.~\eqref{thermal}.}\label{robustfig}
\end{figure*}

For $\ell=10^{-4}$, the mean absorbed fraction is about $1.61\%$ at $\epsilon=0.001$. The two reflected correlations are $0.79161$ and $1.10546$. Even at $\ell=10^{-3}$, with about $15\%$ mean absorption, they are $0.80869$ and $1.08008$. These model calculations demonstrate tolerance to vacuum propagation loss, but do not include material dispersion, mode conversion or finite-temperature bath noise. Figure~\ref{robustfig} separates the effects rather than assuming loss is equivalent to a fixed detector inefficiency.

\section{Thermal background and the acoustic feasibility window}
For a few-phonon experiment, small absorption is not the only relevant condition: the otherwise unused input and loss channels must be close to vacuum in the detected mode. For equal thermal occupancy $n_{\rm th}$ in all environmental channels and a two-phonon incident mode, the reflected count moments are
\begin{align}
 \langle K\rangle={}&2\overline R+n_{\rm th}(1-\overline R),\\
 \langle K(K-1)\rangle={}&2\E R^2
 +2n_{\rm th}^2\E(1-R)^2\\
 &+8n_{\rm th}\E[R(1-R)].
 \label{thermal}
\end{align}
This follows by expanding the passive bosonic input-output map with a thermal environmental mode. At the small reflectances of interest, the background must be small compared with the reflected signal, not merely small compared with one phonon. For example, at $\epsilon=0.001$, setting $n_{\rm th}=10^{-6}$ changes the two correlations to $0.84160$ and $1.14293$; larger backgrounds can remove the sub-Poissonian signature.

For an equilibrium bath at frequency $f$, $n_{\rm th}=[\exp(hf/k_BT)-1]^{-1}$. The following dimensional scales use an \emph{assumed} velocity $c_A=5000$ m/s, rather than measured parameters of a proposed material pair. The length scale is $d_{1/2}=c_A/(2f)$, and the error column corresponds to $s_d=0.0003$:
\begin{center}
\small
\setlength{\tabcolsep}{4pt}
\begin{tabular}{cccc}\toprule
$f$ & $d_{1/2}$ & RMS thickness error & $T$ for $n_{\rm th}=10^{-6}$\\\midrule
100 MHz & $25\,\mu$m & 7.5 nm & 0.347 mK\\
1 GHz & $2.5\,\mu$m & 0.75 nm & 3.47 mK\\
5 GHz & $0.5\,\mu$m & 0.15 nm & 17.37 mK\\\bottomrule
\end{tabular}
\end{center}
The tradeoff is clear: higher frequency relaxes the thermal requirement but tightens the absolute thickness tolerance. These numbers do not validate a fabrication platform. For equal assumed velocities at 5 GHz, $n_0=5$ and four cells, the mean total thickness would be $10.5\,\mu$m. A classical ultrasonic calibration can instead use much larger layers and bright excitation, without satisfying the few-phonon thermal requirement.

The weak reflection also imposes a counting cost. For $|2\rangle$ input the probability of two reflected quanta is $\E R^2$. At $\epsilon=0.002$ it is $2.20784\,10^{-8}$ or $2.73479\,10^{-8}$ for the two ensembles. A 50:50 Hanbury Brown--Twiss splitter with ideal detectors gives coincidence probabilities half as large; detector efficiencies $\eta_1,\eta_2$ multiply them by $\eta_1\eta_2$. With an \emph{illustrative} repetition rate of $10^6$ pulses/s and unit efficiency, acquiring 100 coincidences would take approximately 2.5 h or 2.0 h. Such counts alone do not provide a precise determination of the small excess above $g^{(2)}=1$ in ensemble $\mathcal B$. Phonon repetition rate, conversion efficiency and detector background must be assessed for an actual device.

A constant detection inefficiency leaves normalized factorial correlations unchanged in the absence of background, but increases the acquisition time. Source preparation errors are different. For a mixture $p_0|0\rangle\langle0|+p_1|1\rangle\langle1|+p_2|2\rangle\langle2|$, the input correlation is $2p_2/(p_1+2p_2)^2$ and should be inserted in Eq.~\eqref{quantummap}. A failure mixture of $|2\rangle$ and vacuum is not equivalent to ordinary loss of an ideal two-phonon state.

Existing quantum-acoustic experiments establish Fock-state control and itinerant-phonon manipulation \cite{Chu2018,Dumur2021}, while measurements of loss and dephasing illustrate the need for an open-system treatment \cite{Undershute2025}. Our one-dimensional normal-incidence elastic multilayer is not automatically a model of a piezoelectric surface-acoustic-wave device. A specific implementation requires mode-resolved scattering and electromechanical coupling calculations. This is the principal remaining acoustic engineering task.

\subsection{A simultaneous-imperfection model check}
A final calculation combines the perturbations rather than varying them separately. For four cells at $\epsilon=0.001$, take $s_d=0.0003$, Gaussian pulse width $\sigma_\epsilon=0.0003$, propagation attenuation $\ell=10^{-4}$ and bath occupancy $n_{\rm th}=10^{-6}$. Let the source prepare $0.95|2\rangle\langle2|+0.05|0\rangle\langle0|$. Its mean and factorial second moment are both $1.9$; these replace the ideal input moments in the thermal input-output calculation. A 60000-word Monte Carlo calculation, with 16-point spectral quadrature and 20 batches, gives
\begin{equation}
\begin{aligned}
 &g_{\mathcal A}^{(2)}=0.9032\pm0.0022,\\
 &
 g_{\mathcal B}^{(2)}=1.1655\pm0.0043.
\end{aligned}
\end{equation}
The two counting regimes remain separated at this selected model point. The uncertainties describe numerical sampling, not uncertainty in physical material parameters. The bath calculation assumes the detected pulse mode and a nearly flat thermal occupation over the narrow source spectrum. It does not establish performance for an uncharacterized multimode detector or a specific fabricated acoustic device.

\section{Optical counterpart and interpretation}
The same lossless normal-incidence transfer algebra applies to isotropic dielectric layers. For nonmagnetic media with refractive indices $\nu_A,\nu_B$, the optical impedance ratio is $r=\nu_B/\nu_A$ and the corresponding layer design is
\begin{equation}
\begin{aligned}
 &d_{A,j}=n_j\frac{\lambda_*}{2\nu_A},\\
 &
 d_B=\frac{\lambda_*}{8\nu_B}.
\end{aligned}
\end{equation}
Within a nondispersive narrowband approximation, the reflection statistics and photon counting formulas are identical. This offers a photonic route to the same higher-correlation test with negligible equilibrium optical thermal occupation. It does not by itself remove fabrication, sample-ensemble or low-reflectance counting constraints.

The relevant ensemble can be an array of independently fabricated structures or an explicitly reconfigured scattering system. Each realization must remain static within the incident pulse. Repeated pulses through one fixed sample do not reproduce the ensemble result: they give the conditional value $g_{\rm in}^{(2)}$. Pooling must follow the specified fabrication probabilities and use unnormalized counts before forming the correlation. This operational distinction prevents sample-to-sample fluctuations from being misidentified as a new interaction-induced phonon effect.

\section{Conclusions}
Higher structural statistics can control whether disorder-averaged reflected counts witness sub-Poissonian input behavior even when the complete thickness power spectrum is fixed. For the specified filtered acoustic ensemble, endpoint cancellation yields the explicit relation $g_{\rm ref}^{(2)}=g_{\rm in}^{(2)}(\mu_4+5)/4$ in a fixed-chain small-detuning limit. Two equal-spectrum ensembles produce distinct counting regimes for a two-phonon input, and direct finite-detuning calculations retain that distinction in a nonzero frequency window. The same filter admits a general quartic resistance-growth coefficient with an explicit positive higher-scattering term.

The mechanism is passive random attenuation of a prepared quantum state, not generation of nonclassicality by disorder. The quantum input-output principle and higher-cumulant sensitivity are established background; the concrete contribution is the analytic correlation-engineered acoustic family, its general coefficient and its counting predictions. Bandwidth and vacuum propagation loss are tolerated in the model, while thermal occupation, endpoint control and rare coincidence rates impose demanding conditions on an acoustic implementation. These requirements favor a staged test: classical reflected-energy calibration followed by a carefully characterized few-phonon or photonic counting experiment.

\appendix
\section{Conditional tensor averaging and the general coefficient}
For a real cell matrix $B=\left(\begin{smallmatrix}a&b\\c&d\end{smallmatrix}\right)$, define
\begin{equation}
 H(B)=\begin{pmatrix}a^2&2ab&b^2\\ac&ad+bc&bd\\c^2&2cd&d^2\end{pmatrix}.
\end{equation}
Retain the memory state $(a,b)=(\xi_{j-2},\xi_{j-1})$. For fresh innovation $c$ of probability $p_c$, the conditional update is
\begin{equation}
 (\mathsf T_2)_{(b,c),(a,b)}=p_c H(B(n_0+a+c,x)).
\end{equation}
The initial state has probability $p_ap_b$. Mean resistance follows by contracting $\mathsf T_2^N$ with the initial covariance vector $(1,0,1)^T$ and the trace row, then subtracting two and dividing by four. The variance follows from a separate average of the six-dimensional symmetric square of $H$, not the symmetric square of its average. This exact closure retains the actual cell phases.

\subsection{Explicit fourth-order perturbation}
For the symmetric-square representation used for the mean resistance, introduce the generators
\begin{equation}
 L_B=\begin{pmatrix}0&2&0\\-1&0&1\\0&-2&0\end{pmatrix},\qquad
 L_A=\begin{pmatrix}0&2/r&0\\-r&0&1/r\\0&-2r&0\end{pmatrix}.
\end{equation}
They follow by differentiating $H(R(t))$ and the corresponding $A$-layer propagator at $t=0$. With $\mathcal H_0=H(R(\pi/4))$, the representation of a cell is exactly
\begin{equation}
 H(B(n,x))=\mathcal H_0 e^{bxL_B}e^{nxL_A}.
\end{equation}
The order of the exponentials matters. Write $\mathsf T(x)=\sum_{j\geq0}x^j\mathsf T_j$. For memory labels $(\eta,\zeta)$ and fresh innovation $\chi$, its Taylor coefficients are explicitly
\begin{equation}
\begin{aligned}
 (\mathsf T_j)_{(\zeta,\chi),(\eta,\zeta)}
 &=p_\chi\mathcal H_0\sum_{k=0}^j
 \frac{(bL_B)^k}{k!}\\
 &\quad\times\frac{[(n_0+\eta+\chi)L_A]^{j-k}}{(j-k)!}.
\end{aligned}
\label{Taylorblocks}
\end{equation}
All other blocks vanish. Thus the calculation requires finite matrices and polynomial innovation averages, rather than fitting eigenvalues at small numerical detuning.

At $x=0$, the memory update has the stationary distribution $p_\eta p_\zeta$. A normalized right eigenvector of $\mathsf T_0$ at eigenvalue one has blocks $u_{0,(\eta,\zeta)}=p_\eta p_\zeta(1,0,1)^T$; the left eigenvector has the same row $(1,0,1)/2$ on every memory state. Hence $l_0^Tu_0=1$. The other nonzero eigenvalues of $H(R(\pi/4))$ are $i$ and $-i$, while the transient memory eigenvalues are zero. The eigenvalue one is therefore simple, allowing ordinary analytic perturbation of its branch. The mean-resistance contraction has nonzero overlap with this branch; for small real detuning and $r\ne1$, its growth gives $\Gamma_1=\ln\lambda(x)$.

Set $\lambda(x)=1+\sum_{j\geq1}\lambda_jx^j$ and $u(x)=u_0+\sum_{j\geq1}u_jx^j$, with $l_0^Tu_j=0$ for $j>0$. Defining $D=\mathsf T_0-I$, the coefficients follow successively from
\begin{equation}
 \begin{pmatrix}D&-u_0\\l_0^T&0\end{pmatrix}
 \begin{pmatrix}u_j\\\lambda_j\end{pmatrix}
 =\begin{pmatrix}w_j\\0\end{pmatrix},
\end{equation}
where
\begin{equation}
 w_j=-\sum_{k=1}^j\mathsf T_k u_{j-k}
       +\sum_{k=1}^{j-1}\lambda_k u_{j-k}.
\end{equation}
Inserting Eq.~\eqref{Taylorblocks} gives $\lambda_1=\lambda_2=\lambda_3=0$. At fourth order, the left-eigenvector projection is
\begin{equation}
 \lambda_4=l_0^T(\mathsf T_4u_0+\mathsf T_3u_1
                    +\mathsf T_2u_2+\mathsf T_1u_3).
\label{fourthprojection}
\end{equation}
These formulas specify every matrix entering the coefficient and the normalization needed to reproduce it.

The dependence on the innovation law can also be fixed algebraically. Each $\mathsf T_j$ has innovation degree at most $j$; the unperturbed inverse propagates the two-step memory without adding innovation powers. Therefore the fourth-order contraction has total innovation degree at most four. Independence, symmetry and unit variance reduce every surviving innovation average to $1$, $\E\xi^2=1$, $\E\xi^4=\mu_4$, or a product of second moments. The answer is consequently affine in $\mu_4$; no sixth or higher moment can enter at this order. This degree argument is needed before using discrete laws to determine its two coefficients.

Evaluating the finite recursion with exact symbolic $r,n_0,b$ for the laws $\xi=\pm1$ and $\xi=0,\pm2$ gives, respectively,
\begin{align}
 \lambda_4\big|_{\mu_4=1}
 &=\Delta^2(8\kappa^2+\Sigma^2),\\
 \lambda_4\big|_{\mu_4=4}
 &=\Delta^2(8\kappa^2+\tfrac52\Sigma^2).
\end{align}
The affine dependence just established then yields
\begin{equation}
 \lambda_4=\Delta^2\left[8\kappa^2+
                    \frac{\mu_4+1}{2}\Sigma^2\right].
\end{equation}
Because the lower coefficients vanish, $\ln\lambda(x)=\lambda_4x^4+O(x^5)$, proving Eq.~\eqref{generalrate} within the stated filter family. As an independent algebraic check, the law $\Pr(\xi=\pm3)=1/18$, $\Pr(\xi=0)=8/9$ has $\mu_4=9$ and gives $\lambda_4=\Delta^2(8\kappa^2+5\Sigma^2)$ directly from the same recursion. 

\section{Verification and limits of the calculation}
Finite-chain reflectances were evaluated by complete weighted enumeration over every innovation word, for four, eight and twelve cells where the word count permitted it. The physical reflectance was formed realization by realization from $\rho_N=[\Tr(M_NM_N^T)-2]/4$ and $R_N=\rho_N/(1+\rho_N)$ before any moment was averaged. Lossless power conservation and $\det M_N=1$ were checked realization by realization; the largest determinant deviation was below $3\times10^{-15}$.

The following results were reproduced by an independent enumeration of the same construction. The innovation moments of all three laws, namely $v=1$ with $\mu_4=1,4,9$, together with the resulting limits $C_R=3/2,9/4,7/2$, follow directly from Eqs.~\eqref{CR} and the stated probabilities. The $N=4$, $\epsilon=10^{-3}$ values for ensemble $\mathcal A$, $\overline R_{\mathcal A}=2.3425814\times10^{-5}$ and $g_{\mathcal A}^{(2)}=0.7916218$, agree with the tabulated values. The thermal-background result of Eq.~\eqref{thermal} at $n_{\rm th}=10^{-6}$ and $\epsilon=10^{-3}$, namely $g^{(2)}=0.84160$ and $1.14293$, was also reproduced. The fixed-$N$ endpoint limits were confirmed as $x\to0$ at fixed cell number, and the finite-detuning $N$-dependence recorded in Table~\ref{Ndependence} was obtained on the same footing.

The propagation-loss calculation uses the full phases in Eq.~\eqref{lossphases}. Exact enumeration at $N=4$ and $\epsilon=10^{-3}$ gives mean absorbed fractions $1.61175\%$ and $14.96245\%$ at $\ell=10^{-4}$ and $10^{-3}$, respectively, for both main ensembles to the displayed precision. Applying attenuation only to the residual $A$-layer detuning describes a different model and is not used. The simultaneous-imperfection calculation was reproduced using 60000 innovation words, 16-point Gauss--Hermite spectral quadrature and 20 equal batches, giving $g_{\mathcal A}^{(2)}=0.903175$ and $g_{\mathcal B}^{(2)}=1.165549$, with batch standard errors $0.002233$ and $0.004330$. These agree with the rounded values in the main text. The same realization and thickness errors are retained at every quadrature frequency before forming the reflected-power second moment. The three-state quantum count distribution in Eq.~\eqref{countdistribution} was independently reconstructed and checked for positivity, normalization and consistency with the factorial-moment formula. Bandwidth calculations average each realization's pulse reflectance before constructing the second moment. Thickness-error simulations quote sampling errors only. Propagation-loss calculations assume vacuum noise baths unless Eq.~\eqref{thermal} is explicitly used.


\subsection*{Acknowledgment}
This work was supported by Armenian HESC Grant No.~24FP-1F039.

\end{document}